\documentclass[aps,reprint,article,prl]{revtex4-1}
\usepackage{amsmath}
\usepackage{dcolumn}
\usepackage{graphicx}
\usepackage{float}
\usepackage{color}
\usepackage{ulem}
\usepackage{multirow}

\begin{document}
\title{Local pairing versus bulk superconductivity intertwined by the charge density wave order in Cs(V$_{1-x}$Ta$_{x}$)$_{3}$Sb$_{5}$}

\author{Jinyulin Li,$^{1,*}$ Qing Li,$^{1,*}$ Jinjin Liu,$^{2,3,*}$ Ying Xiang,$^1$ Huan Yang,$^{1,\dag}$ Zhiwei Wang,$^{2,3,4,\ddagger}$ Yugui Yao,$^{2,3}$ and Hai-Hu Wen$^{1,\S}$}

\affiliation{$^1$ National Laboratory of Solid State Microstructures and Department of Physics, Collaborative Innovation Center of Advanced Microstructures, Nanjing University, Nanjing 210093, China}

\affiliation{$^2$ Key Laboratory of Advanced Optoelectronic Quantum Architecture and Measurement, Ministry of Education, School of Physics, Beijing Institute of Technology, Beijing 100081, China}

\affiliation{$^3$ Beijing Key Lab of Nanophotonics and Ultrafine Optoelectronic Systems, Beijing Institute of Technology, Beijing 100081, China}

\affiliation{$^4$ Material Science Center, Yangtze Delta Region Academy of Beijing Institute of Technology, Jiaxing, 314011, China}

\begin{abstract}
There is a common belief that superconductivity and charge density wave (CDW) order accommodate homogenously in real space but compete with each other for the effective density of states in momentum space in CDW superconductors. By measuring resistivity along the $c$-axis in Cs(V$_{1-x}$Ta$_{x}$)$_{3}$Sb$_{5}$, we observe strong superconducting fluctuation behavior coexisting with the CDW order in the pristine CsV$_{3}$Sb$_{5}$, and the fluctuation region becomes narrowed when the Ta doping suppresses the CDW order. The onset transition temperature barely changes with the Ta doping. Therefore, the bulk superconductivity may be established by a doping-independent local pairing, and it can be suppressed in some regions by the spatially variable CDW order along the $c$-axis. Our results violate the above-mentioned belief about CDW superconductors and demonstrate the intricate interaction between superconductivity and CDW order in this kagome superconductor.
\end{abstract}

\maketitle

\section{Introduction}
Charge density wave (CDW) is a periodic modulation of conduction electron density in real space accompanied by lattice distortion in a CDW material. In the momentum space of a low-dimensional system, the CDW is due to the instability of the Fermi surface \cite{CDWreview} primarily via the nesting effect. Based on this picture, the CDW is similar to conventional superconductivity, exhibiting a gap opening on the Fermi surface. Therefore, it is commonly believed that the two orders compete for the same density of states (DOS) near the Fermi energy in momentum space in a CDW superconductor. A piece of evidence for the competition is that the superconducting (SC) transition temperature ($T_\mathrm{c}$) usually increases when the CDW transition temperature ($T_\mathrm{CDW}$) decreases \cite{NbX2,LuIrSi,CuTiSe2,TbTe3,TaSeS,cupratePD}.

$A$V$_{3}$Sb$_{5}$ ($A$ = K, Rb, Cs) is a new family of kagome materials in which the CDW and superconductivity coexist \cite{CsSC,KSC,RbSC}. The period is $2a_{0}\times2a_{0}$ for the 3$Q$ CDW order in the V-Sb layer, with $a_{0}$ as the lattice constant \cite{IlijaNature,GaoHJNature,ChiralCDW,SatoARPES,IlijaRSB,LiHZSP}. The origin of the CDW order in $A$V$_{3}$Sb$_{5}$ is still under debate, that whether it is driven electronically or by the electron-phonon coupling \cite{CDWDaiYM1,CDWtheory,theoryBalents,HardXRay,CDWZhouXJ,CDWnesting,CDWDaiYM2}. Nevertheless, the in-plane CDW order shows obvious anisotropic intensities along three in-plane crystal axes \cite{IlijaNature,GaoHJNature,ChiralCDW,SatoARPES,IlijaRSB,LiHZSP,STMWangZY,IliyaNP}. Meanwhile, in the normal state, the nematic electronic state can be observed in the $ab$-plane \cite{NematicXiangY,WuTNature}, and the in-plane anisotropy decreases simultaneously with the increase of temperature and finally disappears below \cite{WuTNature} or near \cite{NematicXiangY} $T_\mathrm{CDW}$. The in-plane anisotropy may have a close relationship with the three-dimensional (3D) CDW order \cite{HardXRay,STMWangZY,QuanOsc,ResonantXRay,NematicXiangY}. Moreover, there are many interesting phenomena \cite{AHEK,AHECs,ChiralCDW,HuJP,NatureTRSB,NatureChiral,Kerr} accompanying the CDW order, and these phenomena are crucial to help understand the electronic properties in this kagome system. On the other hand, superconductivity in $A$V$_{3}$Sb$_{5}$ is also very interesting. Different experiments prove that the SC gaps are nodeless and probably with anisotropy \cite{CDWnesting,STMFengDL,TDO,uSR,irradiation}. In addition, superconductivity is a strong-coupling one \cite{GaoHJNature,LuXPC,RenCPC} and possibly unconventional \cite{theoryYanBH}. It should be noted that the SC transition width (0.7 - 1.0 K) is always broad compared to the low $T_\mathrm{c}$ ($\approx3.5$ - 4.0 K) in the pristine sample \cite{CsSC,NematicXiangY,AHECs,QuanOscLeiHC,STMYaoYG,AThPower,TidopedChenJG}, and a kink is usually seen in the middle of the resistance transition \cite{CsSC,NematicXiangY,AHECs,QuanOscLeiHC,STMYaoYG}.

In $A$V$_{3}$Sb$_{5}$, the CDW order has a complex, competitive relationship to superconductivity. Under pressure, $T_\mathrm{CDW}$ decreases gradually in CsV$_{3}$Sb$_{5}$, but pressure-dependent $T_\mathrm{c}$ shows an unusual double-dome structure instead of a monotonic increase before the CDW order is suppressed entirely \cite{irradiation,pressureChenXH,pressureChengJG,pressureNature}. The CDW can also be suppressed by chemical doping, and superconductivity can be enhanced  by the Sn \cite{Sndoped}, Ti \cite{Tidoped}, Nb \cite{Nbdoped}, or Ta \cite{Tadoped,GapdopedNature,TadopedXRD} doping. The double-dome feature of $T_\mathrm{c}$ has also been observed in the doping phase diagram versus the Sn doping \cite{Sndoped} or the Ti doping \cite{Tidoped}, while the CDW order is also entirely suppressed by the chemical doping before the second dome of superconductivity is reached at higher doping levels. Distinctive from Ti- or Sn-doped cases which introduces holes into the system and lowers the Fermi level, the Ta doping is supposed to be an isovalent doping scheme. However, the angle-resolved photoemission measurements \cite{Tadoped0.13} reveal that the Ta doping shifts the electron-like band around the $\Gamma$ point towards a slightly deeper binding energy. Meanwhile, the doping lifts the van Hove singularity exactly to the Fermi level in the sample with the highest doping level, which may be the reason for the highest $T_\mathrm{c}$ \cite{Tadoped0.13}.

In this Letter, we present the study on the evolution of the CDW and the SC transition in the Cs(V$_{1-x}$Ta$_{x}$)$_{3}$Sb$_{5}$ system by using the $c$-axis ($\rho_{c}$) and in-plane ($\rho_{ab}$) resistivity measurements. A SC phase with precursor superconductivity can be observed in $\rho_{c}(T)$ curves. The onset transition temperatures $T_{\mathrm{c},\|c}^\mathrm{onset}$ are much higher than those obtained in $\rho_{ab}(T)$ curves, and are nearly doping-independent. The $\rho_{c}(T)$ curves in pristine and Ta-doped samples show a two-step SC transition under a strong magnetic field. This demonstrates a complex interplay between superconductivity and CDW order along the $c$-axis.

\section{Experimental method}
High-quality single crystals of the pristine and the Ta-doped CsV$_{3}$Sb$_{5}$ were grown by a self-flux method with a Cs-Sb binary eutectic mixture as the flux \cite{CsSC,GapdopedNature}. The chemical doping level is confirmed by the energy dispersive x-ray spectrum in scanning electron microscopy (Phenom Prox). The highest doping level we can achieve is about $x_\mathrm{max}\approx 0.14$ in Cs(V$_{1-x}$Ta$_{x}$)$_{3}$Sb$_{5}$. However, the sample with $x_\mathrm{max}\approx0.14$ is too small to carry out the $c$-axis resistance measurements, and the highest doping level in the phase diagram is 0.12. Resistance measurements were performed in a physical property measurement system (PPMS, Quantum Design). Samples were cleaved before the measurements. The in-plane resistivity was measured using a standard four-probe method. Since the thickness of the single crystals is 0.04 - 0.02 mm, it is impossible to measure $c$-axis resistance using a standard four-probe configuration, and it was measured using a four-probe method with a Cobino-shape-like configuration \cite{NematicXiangY,DuG}.

\section{Results}

\subsection{SC and CDW transitions}

\begin{figure}
\centering
\includegraphics[width=8.6cm]{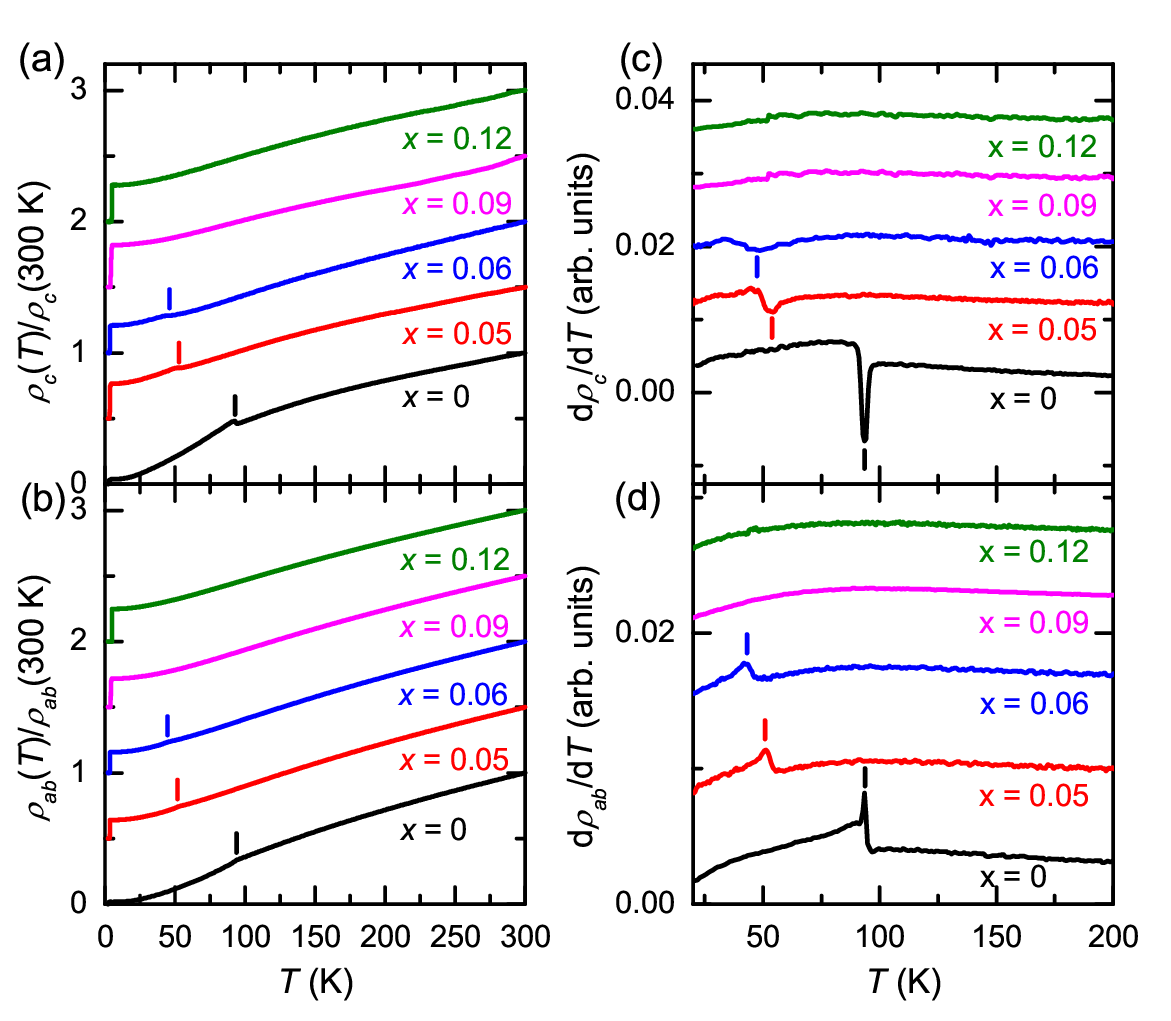}
\caption{Temperature dependence of the (a) $c$-axis and (b) in-plane normalized resistivity with temperature up to 300 K. (c,d) Temperature dependence of (c) $\mathrm{d}\rho_{c}/\mathrm{d}T$ and (d) $\mathrm{d}\rho_{ab}/\mathrm{d}T$ obtained from the data in (a) and (b), respectively. All curves are shifted vertically for clarity. The CDW transitions are indicated by vertical bars.
} \label{fig1}
\end{figure}

Figures~\ref{fig1}(a) and \ref{fig1}(b) show the temperature dependence of the normalized resistivity measured in Cs(V$_{1-x}$Ta$_{x}$)$_{3}$Sb$_{5}$. The residual resistance ratio RRR is defined as $\rho(300\;\mathrm{K})/\rho(7\;\mathrm{K})$ that the value at 7 K represents the normal-state resistivity just above $T_\mathrm{c}^\mathrm{onset}$ for both $c$-axis and $ab$-plane measurements. Since the normal-state resistivity of $\rho_{c}$ or $\rho_{ab}$ is nearly temperature-independent at temperatures up to 7 K, $\rho(7\;\mathrm{K})$ equals to normal-state resistivity at 0 K approximately. The RRR of $\rho_{ab}$ is about 59 in the pristine sample, indicating the high quality of the single crystal. After the substitution by Ta, this value decreases and is about 4 in the sample with $x=0.12$. In the pristine sample, an apparent upward jump can be seen in the $\rho_{c}(T)$ curve with a decrease in temperature when crossing the CDW transition, which is different from a direct drop in the $\rho_{ab}(T)$ curve \cite{NematicXiangY}. Such difference has also been observed in a sister compound of RbV$_{3}$Sb$_{5}$ \cite{RbSC}. The difference is much more apparent in the differential curves shown in Figs.~\ref{fig1}(c) and \ref{fig1}(d); it can also be seen in the resistivity or differential curves measured in Ta-doped samples with $x=0.05$ or 0.06. The repeatable result and the regular evolution of this behavior in doped samples suggest an intrinsic property of the difference. In any case, $T_\mathrm{CDW}$ decreases significantly with the increase of $x$, and the CDW transition finally disappears in samples with $x>0.08$ \cite{GapdopedNature}. The values of $T_\mathrm{CDW}$ are determined from the dip position in the $\mathrm{d}\rho_{c}/\mathrm{d}T$ curves or the peak position in the $\mathrm{d}\rho_{ab}/\mathrm{d}T$ curves, which are indicated by the vertical bars in Fig.~\ref{fig1}.

\begin{figure*}
\centering
\includegraphics[width=17cm]{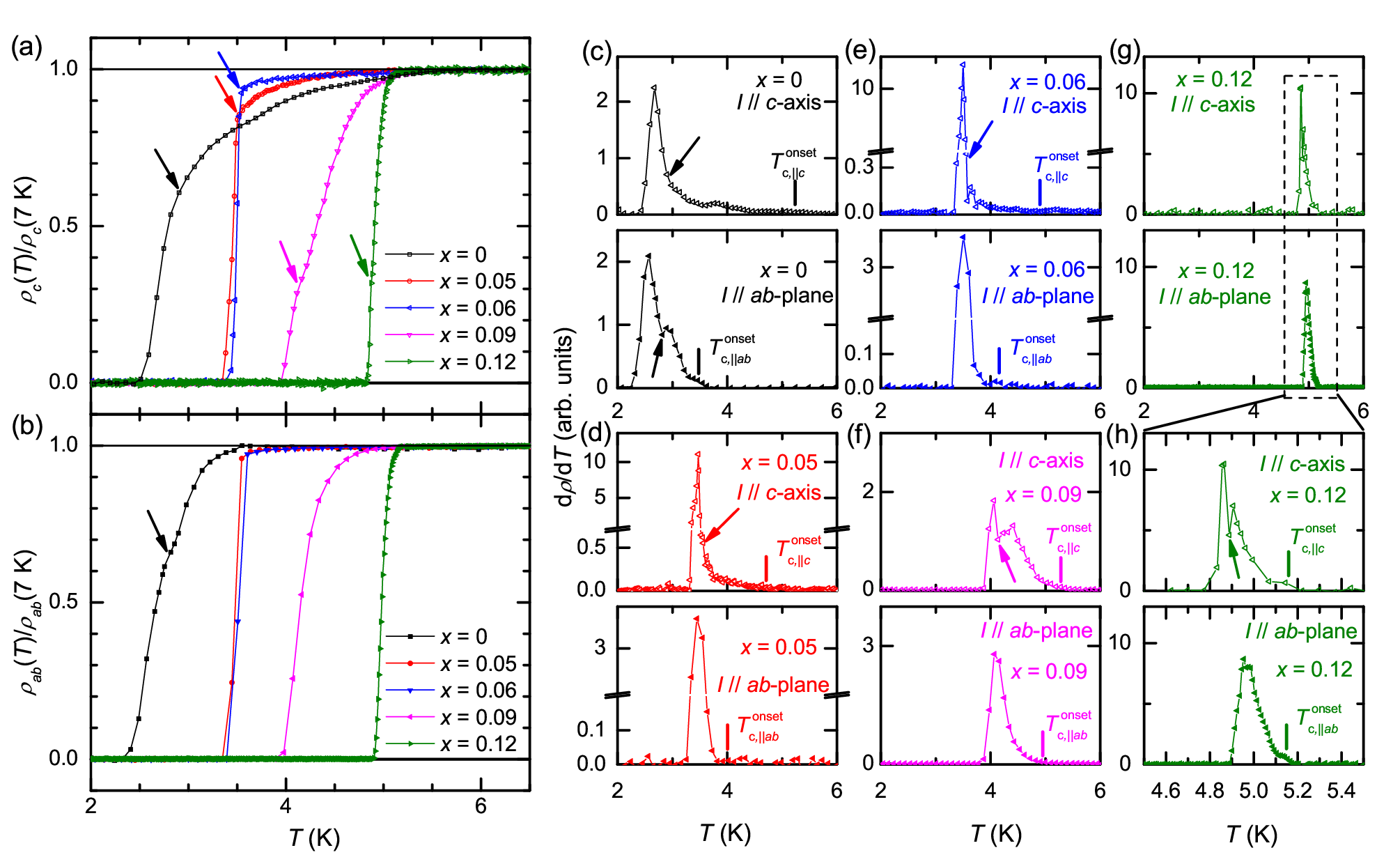}
\caption{Superconducting transitions characterized by (a) the $c$-axis and (b) the in-plane resistivity measurements. Each resistivity curve is normalized by the normal-state resistivity measured at 7 K. Arrows in (a) and (b) indicate the turning points of the two-stage SC transition schematically. (c)-(h) Differential results of $c$-axis (upper panel) and in-plane (lower panel) resistivity for each doping, derived from (a) and (b). In samples with $x=0.09$ (f) and 0.12 (g,h), the two-stage transition is more apparent in $\mathrm{d}\rho_{c}/\mathrm{d}T$ than in $\rho_{c}(T)$. (h) is an enlarged view of SC transition from (g).}
\label{fig2}
\end{figure*}

Accompanied by the suppression of the CDW order, the bulk superconductivity is enhanced by the Ta doping. This conclusion is evident from the temperature-dependent resistivity curves measured near the SC transition (Figs.~\ref{fig2}(a) and \ref{fig2}(b)). The normal-state resistivity $\rho_\mathrm{n}$ of $\rho_{c}$ or $\rho_{ab}$ is nearly temperature independent at temperatures up to 7 K, and the criterion of 99\%$\rho_\mathrm{n}$ is specially selected in order to underline the weak SC-fluctuation phenomenon, and the criterion of 1\%$\rho_\mathrm{n}$ is selected to determine zero-resistance transition temperature because the value is near the resistance measuring accuracy. In the pristine CsV$_{3}$Sb$_{5}$, the zero-resistance temperature $T_\mathrm{c0}$, which is determined by the $1\%\rho_\mathrm{n}$ criterion, is almost the same ($\approx$ 2.5 K) as derived from both $\rho_{c}(T)$ and $\rho_{ab}(T)$ curves. However, the onset transition temperature $T_{\mathrm{c},\|c}^\mathrm{onset}$, determined by the $99\%\rho_\mathrm{n}$ criterion, is 5.2 K derived from the $\rho_{c}(T)$ curve, about twice $T_\mathrm{c0}$ and much higher than $T_{\mathrm{c},\|ab}^\mathrm{onset}\approx 3.4$ K determined in the $\rho_{ab}(T)$ curve. In a conventional superconductor, the thermal-induced superconducting fluctuation range $\delta T_\mathrm{c}\ll T_\mathrm{c}$. Here in CsV$_{3}$Sb$_{5}$, the SC fluctuation along the $c$-axis is very strong, yielding $(T_{\mathrm{c},\parallel c}^\mathrm{onset}-T_{\mathrm{c},\parallel ab}^\mathrm{onset})/T_{\mathrm{c},\parallel ab}^\mathrm{onset}$ of about 50\%. Also, from the $\mathrm{d}\rho_{c}/\mathrm{d}T$ curves in Fig.~\ref{fig2}(c), one can see the non-zero slope of $\rho_{c}(T)$ extends to a much higher temperature than that of $\mathrm{d}\rho_{ab}/\mathrm{d}T$. Therefore, a broad SC-fluctuation-like behavior can be seen in the $\rho_{c}(T)$ curve. In addition, the SC transition is divided into two stages. The black arrow in Fig.~\ref{fig2}(a) marks the turning point of the different SC transition stages schematically, but it is impossible to get the exact diversion temperatures. At the high-temperature stage, $\rho_{c}$ deviates from the normal-state resistivity very smoothly; thus, the SC transition acts like an SC fluctuation. In contrast, at the low-temperature stage, the SC transition becomes sharp as the long-ranged coherence is achieved in the bulk. It should be noted that the kinked feature has also been observed in previous measurements on $\rho_{ab}(T)$ by different groups \cite{CsSC,AHECs,QuanOscLeiHC,STMYaoYG}, which suggests that it is an intrinsic feature in CsV$_{3}$Sb$_{5}$. In addition, a very weak diamagnetic signal of about 0.002\% in volume at 1 T can be observed in CsV$_{3}$Sb$_{5}$ at 3 K by a careful magnetization measurement, and the temperature range of the diamagnetic effect up to about 5 K is consistent with the range in the $\rho_{c}(T)$ curve. This further confirms the SC fluctuation in the sample.

In Ta-doped samples, $T_\mathrm{c0}$ derived from the $\rho_{ab}(T)$ curve is firstly enhanced and reaches a plateau with the increase of the Ta doping level $x$ when $x\leq0.08$, followed by a persistent enhancement to the highest doping level $x \approx 0.14$ \cite{GapdopedNature,TadopedXRD}. This feature can be observed in both $\rho_{c}(T)$ and $\rho_{ab}(T)$ curves shown in Figs.~\ref{fig2}(a) and \ref{fig2}(b), respectively. The $T_{\mathrm{c},\|ab}^\mathrm{onset}$, determined from $\rho_{ab}(T)$ curves, has a similar doping-level dependent trend as $T_\mathrm{c0}$, that nearly monotonically increases from 3.4 to 5.2 K. However, the two-stage transition can only be seen in $\rho_{c}(T)$ curves measured in all samples. The two-stage transition is blurry in the $\rho_{c}(T)$ curves in samples with $x=0.09$ and 0.12, but it is more apparent in $\mathrm{d}\rho_{c}/\mathrm{d}T$ curves shown in Figs.~\ref{fig2}(f)-(h). Meanwhile, the onset transition occurs at a relatively high temperature from 4.7 to 5.3 K for all samples based on $\rho_{c}(T)$ curves. This differs from the behavior of the onset transition in $\rho_{ab}(T)$ curves, in which SC transitions are always sharper than those in $\rho_{c}(T)$ curves.

\subsection{Two-stage SC transition}

\begin{figure}
\centering
\includegraphics[width=8.6cm]{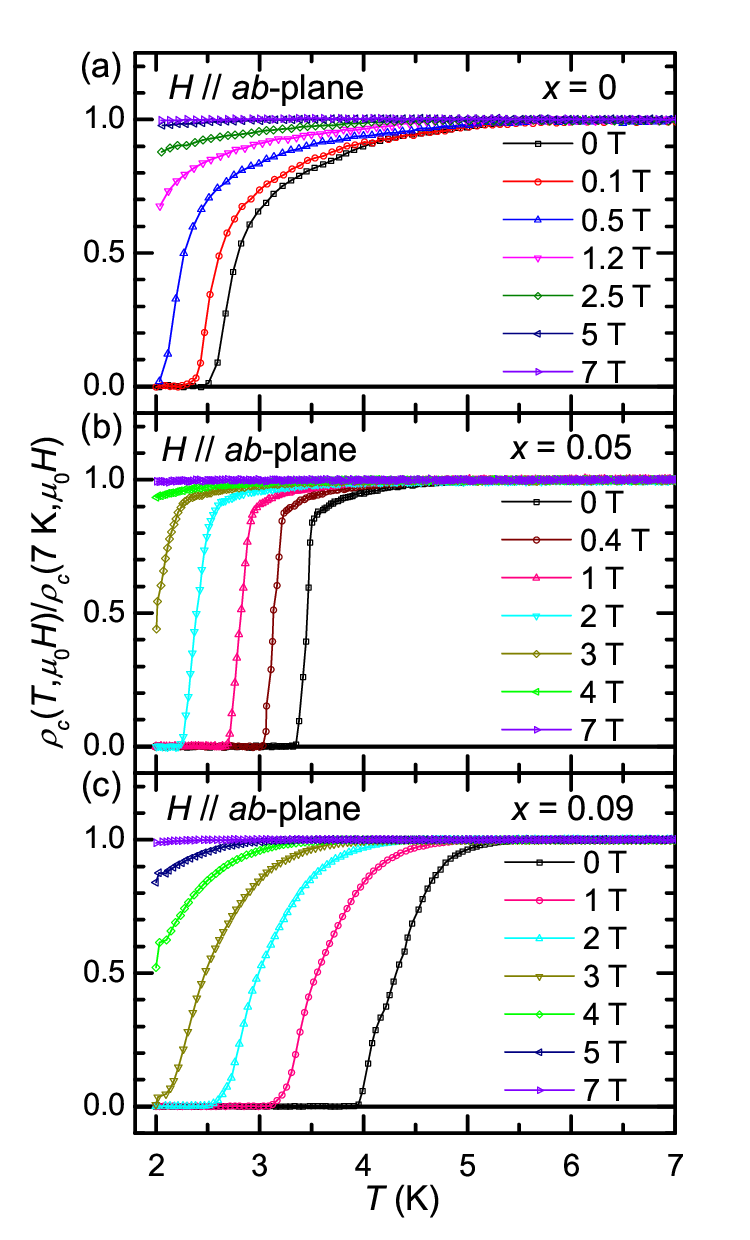}
\caption{Temperature dependence of $c$-axis resistivity measured under magnetic fields in samples with (a) $x=0$, (b) $x=0.05$ and (c) $x=0.09$. The magnetic field is applied parallel to the $ab$-plane. All curves are normalized by the normal-state resistivity at 7 K.
} \label{fig3}
\end{figure}

As mentioned above, the SC transition behaves as a two-stage feature in the pristine and Ta-doped CsV$_{3}$Sb$_{5}$, which is more evident in the resistivity curves measured under magnetic fields. Figure~\ref{fig3} shows $\rho_{c}(T)$ curves measured under different fields and in Cs(V$_{1-x}$Ta$_{x}$)$_{3}$Sb$_{5}$ with $x=0$, 0.05 and 0.09. Again, the SC transition is divided into two stages, i.e., a low-temperature sharp-transition stage and a high-temperature slow-transition stage. Under a magnetic field, the SC transition in the low-temperature stage shifts quickly toward a lower temperature. For example in the pristine sample, a magnetic field of about 1.2 T is strong enough to suppress the sharp transition stage at 2 K. Meanwhile, $\rho_{c}(T)$ curves in this low-temperature stage evolve almost parallelly at different fields. However, the high-temperature stage has a very different behavior under the magnetic field, i.e., the onset transition temperature shifts much more slowly than the zero-resistance temperature. At 2 K, this high-temperature stage is completely suppressed under a magnetic field stronger than 5 T. This observation is also present in the sample with $x=0.05$ and 0.09.

\begin{figure}
\centering
\includegraphics[width=8.6cm]{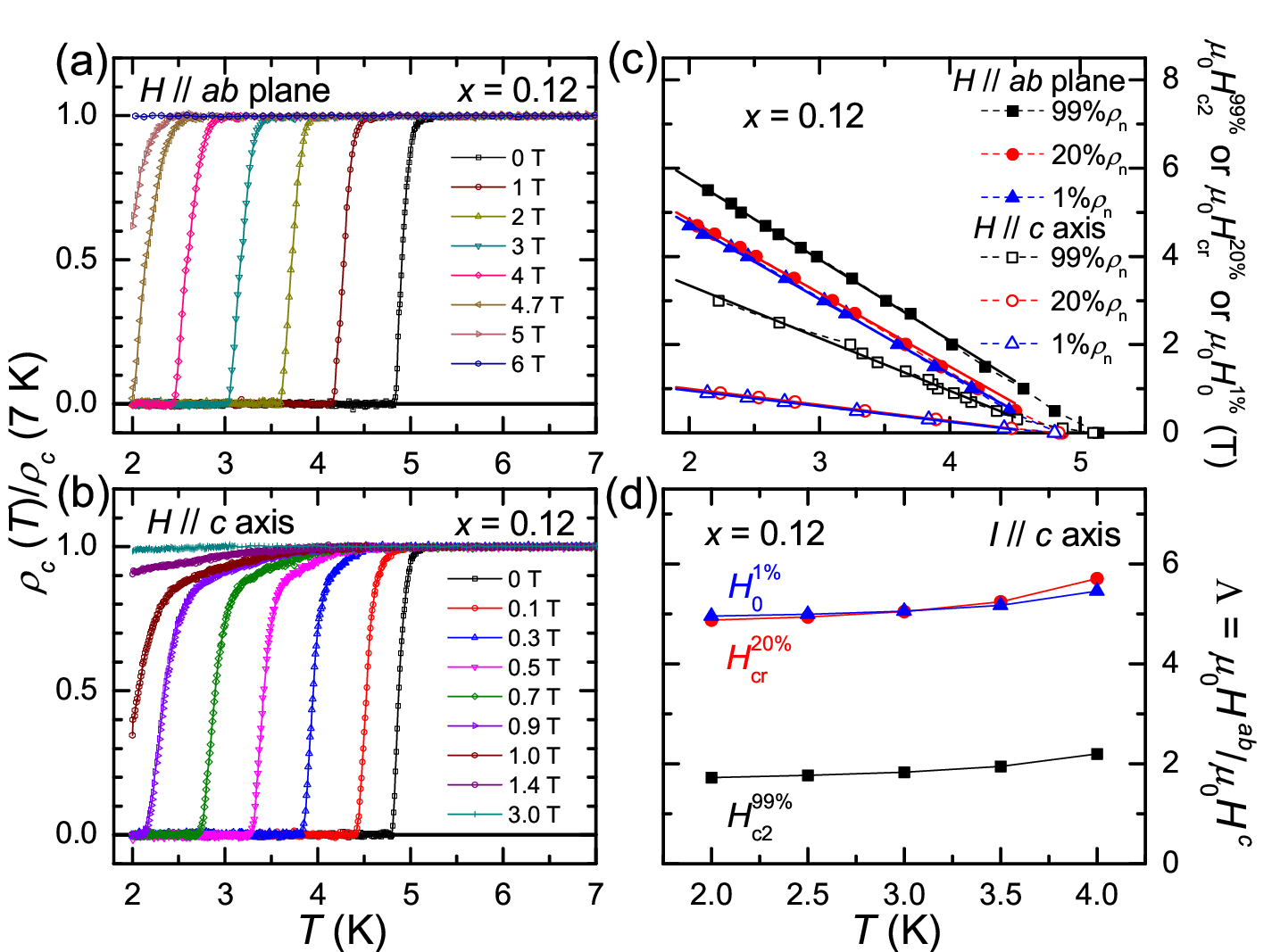}
\caption{(a,b) Temperature dependence of $c$-axis resistivity in Cs(V$_{0.88}$Ta$_{0.12}$)$_{3}$Sb$_{5}$ measured under (a) in-plane and (b) out-of-plane magnetic fields. The two-stage transition is clearer under the out-of-plane magnetic field. (c) Temperature-dependent critical fields determined by 99\%$\rho_\mathrm{n}$, 20\%$\rho_\mathrm{n}$ and 1\%$\rho_\mathrm{n}$ criteria based on the $\rho_{c}(T)$ curves in (a) and (b). The solid lines are linear fittings to the temperature-dependent characteristic transition fields. (d) Out-of-plane anisotropy of different characteristic transition fields at different temperatures.
} \label{fig4}
\end{figure}

Different behaviors of the two SC-transition stages under magnetic fields ($H\parallel ab$-plane) are unclear in the sample with $x=0.12$ (Fig.~\ref{fig4}(a)) because the transition is very sharp. However, one can see that in Fig.~\ref{fig4}(b), the two-stage SC transition is more apparent when the magnetic field is applied along the $c$-axis. Using the criteria of 99\%$\rho_\mathrm{n}$, 20\%$\rho_\mathrm{n}$ and 1\%$\rho_\mathrm{n}$, the critical fields $\mu_{0}H_\mathrm{c2}^{99\%}$, $\mu_{0}H_\mathrm{cr}^{20\%}$ and $\mu_{0}H_{0}^{1\%}$ can be determined from Figs.~\ref{fig4}(a) and \ref{fig4}(b). The characteristic transition fields are plotted in Fig.~\ref{fig4}(c). Under the in-plane magnetic field, $T_{\mathrm{c}}^\mathrm{onset}$ and $T_\mathrm{c0}$ have similar magnetic-field-dependent behavior; while under out-of-plane magnetic field, $T_\mathrm{c0}$ decreases faster than $T_{\mathrm{c}}^\mathrm{onset}$ does. Furthermore, the out-of-plane anisotropy parameter can be obtained from $\Lambda=\mu_{0}H^{ab}/\mu_{0}H^{c}$, where $\mu_{0}H^{ab}$ is the characteristic transition field under the in-plane field and $\mu_{0}H^{c}$ is the the characteristic transition field under out-of-plane field. Different criteria in Fig.~\ref{fig4}(d) stand for different stages of SC transition. Since the resistivity curves under magnetic fields are parallel to each other for the low-temperature stage, values of $\Lambda$ are almost the same for the critical fields with the criterion of 20\%$\rho_\mathrm{n}$ or 1\%$\rho_\mathrm{n}$. However, $\Lambda$ for $\mu_{0}H_\mathrm{c2}^{99\%}$ is smaller than $\Lambda$ for $\mu_{0}H_{0}^{1\%}$, which means in the high-temperature fluctuation stage, superconductivity is less anisotropic. This result strongly suggests that the strong local pairing exists in some regions along $c$-axis.

In the pristine CsV$_{3}$Sb$_{5}$, there is a twofold symmetry of superconductivity in the SC state according to the measurement of $\rho_{c}$ with the in-plane rotation of a magnetic field \cite{NematicXiangY}. It is interesting to investigate whether nematic superconductivity exists in the Ta-doped samples. We do observe a tiny in-plane anisotropy in all samples with the doping level up to $x=0.12$ when the magnetic field rotates in the $ab$-plane. However, the in-plane anisotropy may be induced by a slight misalignment between the current direction and the $c$-axis in the $\rho_{c}$ measurement by the Cobino-shape-like configuration. In the pristine sample, this possibility is ruled out by the anti-phase oscillation of the in-plane magnetoresistance. The magnetoresistance of $\rho_{c}$ is enormous in the pristine sample, i.e., more than 200\% at 2 K and 7 T \cite{NematicXiangY}. However, the magnetoresistance decreases dramatically in the Ta-doped sample, and the value is only just several percent in the Ta-doped sample with $x=0.05$ and negligible in samples with $x\geq 0.06$. Therefore, it is difficult to determine whether the anisotropy is induced by the intrinsic property or the misalignment of electrodes in Ta-doped samples. In Figs.~\ref{fig3} and \ref{fig4}(a), the $\rho_{c}(T)$ curves are measured with the magnetic field roughly along the direction with the maximal in-plane upper critical field. The $\rho_{c}(T)$ curves measured with magnetic fields along other in-plane directions also show the two-stage superconducting transition.

\subsection{Phase diagram with Ta doping}

\begin{figure}
\centering
\includegraphics[width=8.6cm]{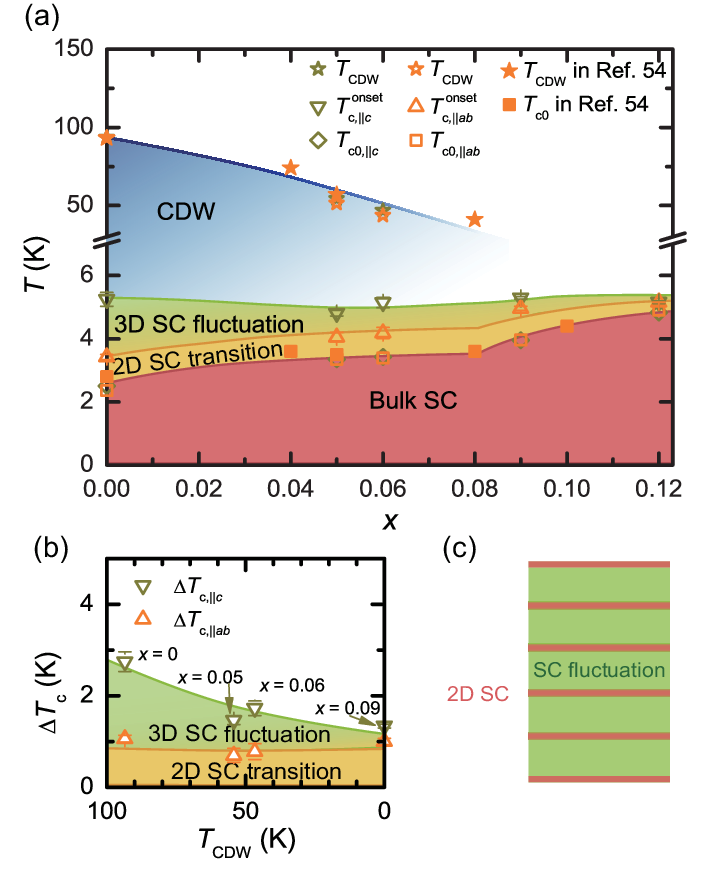}
\caption{(a) Phase diagram of Cs(V$_{1-x}$Ta$_{x}$)$_{3}$Sb$_{5}$ with Ta doping. Values are almost identical for $T_\mathrm{CDW}$ or $T_\mathrm{c0}$ determined by the $c$-axis and the $ab$-plane resistive measurements. The data shown by solid symbol is taken from Ref. \cite{GapdopedNature}. $T_\mathrm{c}^\mathrm{onset}$ shows a discrepancy between the two different measurements. (b) Relationship between superconducting transition width and $T_\mathrm{CDW}$. The difference between these two transition width of $\Delta T_{\mathrm{c},\parallel c}-\Delta T_{\mathrm{c},\parallel ab}$ approximately equals to the SC fluctuation temperature region. Error bars here is same as error bars in $T_\mathrm{c}^\mathrm{onset}$ as the error in $T_\mathrm{c0}$ is negligible. (c) Schematic image of the real-space separation of the 3D SC fluctuation (light green) and the 2D superconducting V$_{3}$Sb$_{2}$ layer (red). The error bars in (a) and (b) are obtained by the corresponding temperature window in the presence of the resistance variation of $0.2\%\rho_\mathrm{n}$.
} \label{fig5}
\end{figure}

In the $c$-axis and $ab$-plane resistive measurements, the values of $T_\mathrm{CDW}$, $T_\mathrm{c0}$, and $T_\mathrm{c}^\mathrm{onset}$ can be obtained. A phase diagram based on these values is plotted in Fig.~\ref{fig5}(a). The values of $T_\mathrm{CDW}$ are almost the same based on the data of $\rho_{c}(T)$ and $\rho_{ab}(T)$, although the CDW transition behaves differently in these curves. $T_\mathrm{CDW}$ decreases with the doping level, and the CDW transition is finally invisible in samples with $x>0.08$ \cite{GapdopedNature}. For the SC transition, the value of $T_\mathrm{c0}$ determined from $\rho_{c}(T)$ and $\rho_{ab}(T)$ is almost the same. Here, $T_\mathrm{c0}$ is first enhanced in the sample with $x=0.05$ when compared to the value of the pristine sample, and then it reaches a plateau at medium doping levels \cite{GapdopedNature} of $x=0.08$. After the plateau, the CDW order is entirely suppressed and $T_\mathrm{c0}$ continues to increase with $x$. The doping-dependent $T_\mathrm{CDW}$ and $T_{\mathrm{c0},\|ab}$ are consistent with the previous reports \cite{GapdopedNature,TadopedXRD}. Here, from our data, $T_{\mathrm{c},\|ab}^\mathrm{onset}$, determined by $\rho_{ab}(T)$ curves, has a similar doping-dependent behavior with $T_{\mathrm{c0},\parallel ab}$ when $x<0.08$, resulting in the nearly constant transition width of $\Delta T_{\mathrm{c},\parallel ab}\equiv T_{\mathrm{c},\parallel ab}^\mathrm{onset}-T_{\mathrm{c0},\parallel ab}$. Meanwhile, there is big difference between $T_{\mathrm{c},\parallel c}^\mathrm{onset}$ and $T_{\mathrm{c},\parallel ab}^\mathrm{onset}$, and the wide temperature gap corresponds to the SC fluctuation along the $c$-axis. The transition width of $\Delta T_{\mathrm{c},\parallel c}\equiv T_{\mathrm{c},\parallel c}^\mathrm{onset}-T_{\mathrm{c0},\parallel c}$ or the temperature gap of the SC fluctuation $T_{\mathrm{c},\parallel c}^\mathrm{onset}-T_{\mathrm{c},\parallel ab}^\mathrm{onset}\simeq \Delta T_{\mathrm{c},\parallel c}-\Delta T_{\mathrm{c},\parallel ab}$ gradually shrinks with the increase of the Ta doping. This effect is clearly shown in Fig.~\ref{fig5}(b), in which one can see a positive correlation between $T_\mathrm{CDW}$ and the temperature gap of the SC fluctuation.

\section{Discussion}
In Cs(V$_{1-x}$Ta$_{x}$)$_{3}$Sb$_{5}$, we observe an interesting SC fluctuation behavior in the $\rho_{c}(T)$ curves, but this behavior is not obvious in the $\rho_{ab}(T)$ curves. The transition width in the in-plane resistivity is almost constant in different samples with different doping levels, which is clearly shown as an almost constant 2D SC transition width in Fig.~\ref{fig5}(b). However, the transition width shows a clear variation in the $\rho_{c}(T)$ curves, that the width decreases with the increase of Ta doping or the decrease of $T_\mathrm{CDW}$. This is due to the obvious two-stage transitions: the onset transition temperature dominated by 3D SC fluctuation is almost unchanged with the varying $x$, while the zero-resistance transition temperature dominated by the bulk superconductivity increases with the increasing $x$. The low-temperature superconducting transition is very sharp, and this part of superconductivity has a lower critical field and a possible 2D character. In contrast, the high-temperature superconducting transition is very broad, and this part of superconductivity has a higher critical field and a possible 3D character. This kind of two-stage SC transition can be observed in all samples, and it should be an intrinsic feature. Even in the sample with $x=0.12$, this transition behavior shows up in the $\rho_{c}(T)$ curves measured at a high magnetic field (Fig.~\ref{fig4}(b)). Then we will discuss the possible origin of this observation.

The round superconducting transitions have been observed and discussed in several iron-based superconductors, such as FeSe \cite{FeSefluctuation} and NaFeAs \cite{NaFeAs}, suggesting a strong SC fluctuation. Usually, strong SC fluctuation is expected in a superconductor with a dilute charge carrier density. Then, the material is near the crossover region from Bardeen-Cooper-Schrieffer (BCS) to Bose-Einstein condensation (BEC) \cite{BCSBEC}. One can use the number of SC electrons in unit coherent volume $V_\mathrm{coh}n_\mathrm{pair}$ to estimate the overlapping SC pairs. Here, the coherent volume $V_\mathrm{coh}=4\pi\xi_{ab}^{2}\xi_{c}/3$, and the density of Cooper pairs $n_\mathrm{pair}$ equals half of the density of SC electrons $n_\mathrm{s}$. In the BCS case, $V_\mathrm{coh}n_\mathrm{pair}\gg1$, while $V_\mathrm{coh}n_\mathrm{pair}\ll1$ in the BEC case \cite{BCSBEC}. For example, the calculated value of $V_\mathrm{coh}n_\mathrm{pair}$ is about 1.0 for Bi$_{2}$Sr$_{2}$CaCu$_{2}$O$_{8}$ (Bi2212), which confirms that this superconductor is in the region near the BCS-BEC crossover \cite{FeSe}. In the pristine CsV$_{3}$Sb$_{5}$, $\mu_{0}H_{\mathrm{c2},ab}\approx5.68$ T and $\mu_{0}H_{\mathrm{c2},c}=0.25$ T at 2 K \cite{NematicXiangY}, and the coherence length can be calculated via $\mu_{0}H_\mathrm{c2}=\Phi_{0}/(2\pi\xi^{2})$. Together with $n_\mathrm{s}=1.1\times10^{21}\;\mathrm{cm}^{-3}$ from Ref.~\cite{Tadoped}, the calculated value of $V_\mathrm{coh}n_\mathrm{pair}$ is about $2.3\times10^{4}$, much larger than 1. It means that CsV$_{3}$Sb$_{5}$ is far away from the BCS-BEC crossover. In addition, the Ginzburg number $Gi$ can be used to characterize the magnitude of the SC fluctuation \cite{FeSe}. Here, $Gi=1.7\times10^{-11}\;T_\mathrm{c}^{2}\kappa^{4}/(\mu_{0}H_\mathrm{c2}\epsilon^{2})$ with the Ginzburg-Landau parameter $\kappa$ and the anisotropy parameter $\epsilon=H_{\mathrm{c2},c}/H_{\mathrm{c2},ab}$. We use $T_\mathrm{c0}=2.5$ K, $\kappa=5.5$ \cite{Tadoped}, $\mu_{0}H_{\mathrm{c2},c}=0.25$ T, $\epsilon=0.044$, and the estimated value of $Gi$ is $2\times10^{-4}$. This value is minimal and comparable with those in conventional superconductors. For example \cite{FeSe}, $Gi$ is about $5\times10^{-3}$ in MgB$_{2}$, while this value is 380 in Bi2212. These calculations demonstrate that the CsV$_{3}$Sb$_{5}$ is far from the BCS-BEC crossover, and the SC fluctuation should not be strong in the material from the conventional understanding.

We note that a pseudogap phase is observed in the pristine CsV$_{3}$Sb$_{5}$ \cite{GaoHJNature}. The pseudogap closes at a temperature above 4.2 K, much higher than the bulk $T_\mathrm{c0}$; the suppression field is also much greater than the bulk $\mu_{0}H_\mathrm{c2}$ derived from the measurements of $\rho_{ab}$. In addition, the enhanced superconductivity is also observed in the experimental data by the point-contact spectroscopy measurement \cite{LuXPC,RenCPC}, and the SC-related zero-bias conductance can extend to about 5 K. The SC-fluctuation-like behavior in this work may be related to these observations in the pristine CsV$_{3}$Sb$_{5}$. From another point of view, the charge-4$e$ and charge-6$e$ superconductivity \cite{charge4e} are observed in CsV$_{3}$Sb$_{5}$. The charge-4$e$ superconductivity is supposed to appear in a nematic superconductor even above $T_\mathrm{c}$ by theoretical proposals \cite{charge4eNematic1,charge4eNematic2}. Therefore, the nematic superconductivity \cite{NematicXiangY} and SC fluctuation above $T_\mathrm{c}$ may be also related to the high-charge superconductivity in CsV$_{3}$Sb$_{5}$. However, evidence of the pseudogap phase or the high-charge superconductivity is lacking in Ta-doped samples, and the origin of the SC fluctuation observed here is still an open question.

Concerning the crystal structure, Cs(V$_{1-x}$Ta$_{x}$)$_{3}$Sb$_{5}$ is a layered kagome metal with a quasi-two-dimensional structure, and the V$_{3}$Sb$_{2}$ layer plays a role as the conducting and the SC plane. The SC fluctuation should be strong in a two-dimensional (2D) system such as SC thin films in which the thickness of the SC layer is much smaller than the coherence length \cite{SCF}. This is the situation in some high-$T_\mathrm{c}$ cuprates with large anisotropy values, and the in-plane excess conductivity induced by the SC fluctuation follows the picture of 2D Aslamazov-Larkin theory \cite{Bi2212,Tl2212,ShortkFSCF}. Then, the bulk superconductivity can be established by the interlayer Josephson coupling of the CuO$_{2}$ planes. In cuprates, the excess conductivity of SC fluctuation can be observed in both in-plane and $c$-axis resistivity curves \cite{Bi2212,Tl2212}. However, in the case of CsV$_{3}$Sb$_{5}$, the precursor superconductivity can only be observed in the $\rho_{c}(T)$ curves, while it is absent in the $\rho_{ab}(T)$ curves. In other words, partial superconductivity occurs along the $c$-axis in this V-based system. This differs from the 2D SC fluctuation in layered compounds such as cuprates, where the SC fluctuation is more substantial in the $ab$-plane than along the $c$-axis \cite{fluctuationBook,LaBa214,LaSr214}. In Cs(V$_{1-x}$Ta$_{x}$)$_{3}$Sb$_{5}$, the current majorly flows in the V$_{3}$Sb$_{2}$ layer when the current flows parallel to the $ab$-plane. When partial superconductivity occurs along the $c$-axis, the $\rho_{c}$ drop may have little influence on $\rho_{ab}$. For example, the ratio of $\rho_{c}/\rho_{ab}$ is about 23 at 7 K in CsV$_{3}$Sb$_{5}$. If the interlayer conductance has isotropic behavior, the 20\% resistivity drop of the interlayer part can only produce a drop of less than 1\% in the $\rho_{ab}$ curve based on the parallel circuit consisting of better-conductive V$_{3}$Sb$_{2}$ layers and worse-conductive interlayer parts. That may be the reason why we cannot see the SC-fluctuation-like behavior in $\rho_{ab}(T)$ curves, and it is similar to the fact that the pseudogap feature is more evident in the measurement of $c$-axis optical conductivity \cite{Optical} than the in-plane optical conductivity. This fact also suggests that the possible local pairing does not happen in V$_{3}$Sb$_{2}$ layers and is likely to behave as a 3D feature assisted by the interlayer coupling or other mechanisms. Since the $c$-axis resistivity is much larger than the $ab$-plane one, the supercurrent of the partial superconductivity contributing to the diamagnetic effect can be easily scattered by other non-superconducting quasiparticles. In addition, the fragile 3D superconductivity can also be interrupted by the non-superconducting V$_{3}$Sb$_{2}$ layers. This may explain the extremely small diamagnetic volume in the SC-fluctuation region. The conclusion is consistent with the anisotropy analysis for the two-stage SC transition. The possible local pairing phase has a much smaller anisotropy (about 1.7 at 2 K) in the high-temperature state than the 2D SC in the low-temperature stage (about 4.9 at 2 K) in the sample with $x=0.12$ (Fig.~\ref{fig4}(d)). Anyhow, the local pairing makes the vanadium-based superconductor similar to the precursor superconductivity in cuprates \cite{cupratePD} although the mechanism of the local pairing may be different from that in cuprates.

The CDW order in the pristine CsV$_{3}$Sb$_{5}$ is a 3D one \cite{HardXRay,STMWangZY,QuanOsc,ResonantXRay}, and the periodic modulation of DOS is also present along the $c$-axis. Based on $\rho_{c}(T)$, the bulk superconductivity occurs at a lower temperature, while the SC fluctuation is strong at higher temperatures. This suggests that CDW order may suppress DOS differently at different positions along the $c$-axis in the pristine sample, and this may induce the periodic SC order parameter along the $c$-axis. With the Ta substitution, the CDW is suppressed, and the 3D CDW changes to a quasi-2D one \cite{TadopedXRD}. Then, the 3D superconductivity is easier to achieve; therefore, the bulk superconductivity is enhanced while the temperature range of the SC fluctuation shrinks. Here, based on our experimental data, the possible local pairing (characterized by $T_{\mathrm{c},\|c}^\mathrm{onset}$) is almost independent of $T_\mathrm{CDW}$ when the doping level of Ta changes, and the SC-fluctuation region has a much smaller anisotropy than that of the bulk superconductivity. We propose a model to explain the observation. It is illustrated as a schematic drawing in the inset of Fig.~\ref{fig5}. In some regions along the $c$-axis, the CDW order is ``weak'' and not detrimental to superconductivity, which leads to the existence of the local pairing. It is this local pairing that induces the onset of SC transition which is weakly dependent on doping. In other regions, the CDW is robust, which suppresses the superconductivity. By doping Ta to the V sites, the CDW is globally suppressed, and the bulk superconductivity can be established at a higher temperature. This picture also interprets why the SC transition is intrinsically broad compared to the low $T_\mathrm{c}$ in the material.

\section{Conclusion}

In summary, we have witnessed a strong SC fluctuation effect in Cs(V$_{1-x}$Ta$_{x}$)$_{3}$Sb$_{5}$ by measuring the $c$-axis resistivity. This is illustrated by an intrinsic feature showing a large temperature gap between the onset and zero-resistance transition temperatures, especially in the pristine sample. The onset transition temperature of the $c$-axis resistivity is hardly influenced by the doping of Ta. In contrast, the zero-resistance transition temperature shows a monotonic increase accompanied by the suppression of the CDW temperature versus doping. Our results suggest a novel competition mechanism between the CDW and superconductivity, which could be induced by the spatially variable CDW order along the $c$-axis. These observations greatly enrich the phase diagram of this family of vanadium-based superconductors and shed new light on understanding the complex interplay between the CDW and superconductivity.

\section*{Acknowlegement}

We acknowledge helpful discussions with Shunli Yu. This work was supported by the National Key R\&D Program of China (Nos. 2022YFA1403201, 2022YFA1403400, and 2020YFA0308800), the National Natural Science Foundation of China (Nos. 11927809, 12061131001, 11974171, 92065109, and 12204231), the Fundamental Research Funds for the Central Universities (No. 020414380208). Z.W. thanks the Analysis \& Testing Center at BIT for assistance in facility support.

$^*$ These authors contributed equally to this work.

Corresponding authors:

$^\dag$ huanyang@nju.edu.cn

$^\ddag$ zhiweiwang@bit.edu.cn

$^\S$ hhwen@nju.edu.cn

\end{document}